\documentclass[manuscript, nonacm]{acmart} 
\AtBeginDocument{%
  }

\usepackage[utf8]{inputenc}
\usepackage{rotating} 
\usepackage{booktabs} 
\usepackage{lscape} 
\usepackage{multirow}  
\usepackage{array}     
\usepackage{afterpage}
\usepackage{pdflscape}
\usepackage{color,xcolor}
\usepackage{hyperref}

\definecolor{FlickrPink}{rgb}{1.0,0.00,0.52}
\definecolor{BabyBlue}{rgb}{0.54,0.81,0.94}

\copyrightyear{2025}
\acmYear{2025}
\setcopyright{none}
\acmBooktitle{none}

\setcopyright{none} 
\renewcommand\footnotetextcopyrightpermission[1]{} 

\ccsdesc[500]{Information systems~Recommender systems}
\ccsdesc[500]{Human-centered computing~User studies}
\ccsdesc[300]{Information systems~Personalization}
\ccsdesc[300]{Human-centered computing~Human computer interaction (HCI)}
\ccsdesc[300]{Computing methodologies~Qualitative studies}
\ccsdesc[200]{Information systems~Evaluation of retrieval results}

\keywords{Experienced Serendipity, Recommender Systems, User Experience, Qualitative Research, Grounded Theory, Content Discovery}

\begin{document}

\title{What Is Serendipity? An Interview Study to Conceptualize Experienced Serendipity in Recommender Systems}

\author{Brett Binst}
\email{brett.binst@vub.be}
\orcid{0000-0002-0243-9952}
\affiliation{%
  \institution{imec-SMIT, Vrije Universiteit Brussel}
  \city{Brussel}
  \country{Belgium}
}

\author{Lien Michiels}
\email{lien.michiels@vub.be}
\orcid{0000-0003-0152-2460}
\affiliation{%
  \institution{imec-SMIT, Vrije Universiteit Brussel}
  \city{Brussel}
  \country{Belgium}
}
\affiliation{%
  \institution{University of Antwerp}
  \city{Antwerp}
  \country{Belgium}
}

\author{Annelien Smets}
\email{annelien.smets@vub.be}
\orcid{0000-0003-4771-7159}
\affiliation{%
  \institution{imec-SMIT, Vrije Universiteit Brussel}
  \city{Brussel}
  \country{Belgium}
}

\begin{abstract}

Serendipity has been associated with numerous benefits in the context of recommender systems, e.g., increased user satisfaction and consumption of long-tail items.
Despite this, serendipity in the context of recommender systems has thus far remained conceptually ambiguous.
This conceptual ambiguity has led to inconsistent operationalizations between studies, making it difficult to compare and synthesize findings. 
In this paper, we conceptualize the user's experience of serendipity. 
To this effect, we interviewed 17 participants and analyzed the data following the grounded theory paradigm.
Based on these interviews, we conceptualize experienced serendipity as \textit{a user experience in which a user unintentionally encounters content that feels fortuitous, refreshing, and enriching}.
We find that all three components---fortuitous, refreshing and enriching---are necessary and together are sufficient to classify a user's experience as serendipitous.
However, these components can be satisfied through a variety of conditions.
{Our conceptualization unifies previous definitions of serendipity within a single framework, resolving inconsistencies by identifying distinct flavors of serendipity. 
It highlights underexposed flavors, offering new insights into how users experience serendipity in the context of recommender systems. 
By clarifying the components and conditions of experienced serendipity in recommender systems, this work can guide the design of recommender systems that stimulate experienced serendipity in their users, and lays the groundwork for developing a standardized operationalization of experienced serendipity in its many flavors, enabling more consistent and comparable evaluations.}

\end{abstract}



\maketitle

\section{Introduction}
Much of the content users encounter online is not actively searched for but recommended. 
Popular platforms like YouTube, Spotify, and Instagram all present users with automatically suggested content. 
At the core of these platforms are recommender systems (RecSys) which analyze user preferences in order to suggest items from vast catalogs aiming to recommend interesting content~\cite{ricci_recommender_2022}.

RecSys can assist users with a variety of tasks~\cite{herlocker_evaluating_2004}. 
For example, they can help users keep up to date with regularly consumed content by suggesting summaries of recent matches in their favorite soccer league. 
In such cases, evaluation metrics focusing on accuracy are appropriate, as they measure how well the recommendations match the users' preferences. 
Conversely, RecSys can also support users in exploring the item space, introducing them to new and unexpected content. 
In these cases, beyond-accuracy metrics become essential, serendipity in particular~\cite{de_gemmis_investigation_2015,ge_beyond_2010,mcnee_being_2006}.
    
Serendipity, in the context of RecSys, captures the idea of desirable exploration within the item space~\cite{de_gemmis_investigation_2015}. 
Several benefits are associated with serendipity, including increased user satisfaction~\cite{pastukhov_offline_2022}, stimulation of long-tail item consumption~\cite{sa_diversity_2022}, support for users in achieving self-actualization~\cite{graus_let_2018,sullivan_reading_2019,knijnenburg_recommender_2016}, and mitigation of the overspecialization problem~\cite{stitini_improved_2022}. 

However, realizing these benefits requires a clear understanding of what serendipity is. 
Despite its potential advantages, the concept of serendipity remains conceptually ambiguous in the context of RecSys~\cite{ziarani_serendipity_2021, kotkov_dark_2024}.
This conceptual ambiguity led researchers to adopt diverse approaches to operationalize the concept~\cite{ziarani_serendipity_2021,kotkov_dark_2024}, resulting in inconsistent metrics of serendipity, which complicate the synthesis of findings across studies~\cite{kotkov_dark_2024}. 
Hence, the conceptual ambiguity of serendipity presents a significant challenge for those who seek to study it~\cite{kotkov_challenges_2016} or design RecSys and platforms that afford it~\cite{smets_serendipity_2022-1}.

While several have tried to address this research gap, e.g.,~\cite{ziarani_serendipity_2021, kotkov_rethinking_2023, smets_serendipity_2022-1, smets_serendipity_2022} as is discussed in greater detail in Section \ref{sec:related}, none have focused their attention on understanding users' subjective experiences of serendipity with RecSys.
Yet, we argue that in order to design RecSys that afford serendipity, or understand what strategic objectives it may serve~\cite{smets_serendipity_2022-1}, we first need a thorough understanding of how users experience serendipity.
Our study fills this gap by conceptualizing experienced serendipity in the context of RecSys.
To achieve this, we adopt a user-centered approach, conducting interviews with 17 participants and analyzing them using grounded theory~\cite{strauss_basics_2015}. 
This method is particularly well-suited for uncovering the nuances of serendipity and has been successfully used in prior work to conceptualize serendipity in other contexts~\cite{makri_coming_2012}. 
{The details of our methodology can be found in Section \ref{sec:meth}}.


Based on our findings, we conceptualize experienced serendipity as \textit{a user experience in which the user unintentionally encounters content that feels fortuitous, refreshing, and enriching}.
These three main components---fortuitous, refreshing, and enriching---serve as necessary and sufficient components for delineating experienced serendipity. 
{Crucially, in Section \ref{sec:results}, we describe our conceptual framework in detail and demonstrate how these main components can be satisfied through different conditions, allowing for a variety of flavors of experienced serendipity, thus both deepening and broadening our understanding of experienced serendipity.}

Our conceptualization has important implications for future research and practice, as we discuss in Section \ref{sec:discussion}.
First, it reconciles apparent contradictions in previous conceptualizations, e.g., the tension between relevance---often operationalized as similarity with the user's profile---and novelty---frequently operationalized as distance from the user's profile.
Second, it uncovers flavors of experienced serendipity that have thus far been underexplored in the literature, e.g., \textit{taste-broadening serendipity}.
Third, it provides additional depth to the concept, thus providing inspiration to RecSys designers to build systems that afford more serendipity. 
Finally, in Section \ref{sec:discussion}, we discuss critical implications for operationalizing experienced serendipity, paving the way for more consistent evaluations of experienced serendipity.

\section{Related Work} \label{sec:related}

\subsection{Conceptualizing Serendipity}

Serendipity is widely considered to be an important aspect of human life~\cite{makri_coming_2012,bandura_psychology_1982,busch_towards_2022,roberts_serendipity_1989}.
Yet, at the same time, it is known as a `slippery concept’~\cite{makri_coming_2012}, that is not fully understood.
In this Section, we highlight a few notable attempts at conceptualizing serendipity that have most influenced this paper. 

In 2012, \citet{makri_coming_2012} described a process model of serendipity{, where an individual makes a new connection due to a mix of unexpected circumstances triggering a moment of insight, which is subsequently exploited, leading to a valuable unanticipated outcome.} 
To arrive at this model, they conducted semi-structured interviews, citing success with this same methodology in previous attempts to conceptualize serendipity.
Notably, they describe taking an `active interview’ approach in which interviewees ``lead with the story of their examples’’, as have we. 
Also similarly, they later applied grounded theory~\cite{strauss_basics_2015} to analyze results. 
 
\citet{bjorneborn_three_2017} takes a design-oriented approach instead and proposes three key affordances to facilitate serendipity in online and offline environments. 
He argues that different environments, e.g., online platforms, can be---deliberately or unconsciously---designed to afford more or less potential for {stimulating} the experience of serendipity.      
His work inspired \citet{smets_serendipity_2022}, who propose an `affordance feature repository’ for serendipity in RecSys, i.e., an overview of features of the online platform that stimulate user experiences of serendipity.
{According to \citet{smets_serendipity_2022}, some platforms and RecSys are more likely to afford serendipity than others, allowing for a greater potential for experiencing serendipity.} 

In a different work, \citet{smets_serendipity_2022-1} contributes to the conceptualization of serendipity by positing it is a value that can be designed for. She distinguishes between three levels of serendipity: 
\begin{itemize}
     \item {\textbf{Intended serendipity: }Why design for serendipity? 
     This level focuses on the strategic goals, such as increasing sales of niche items.}
     \item {\textbf{Afforded serendipity: }How can we stimulate experienced serendipity? 
     This level deals with the capacities of the system to afford experienced serendipity.}
     \item {\textbf{Experienced serendipity: }What characterizes experienced serendipity? 
     This level focuses on how to evaluate serendipity as a \textbf{user experience}.}
 \end{itemize}

Following \citeauthor{smets_serendipity_2022-1} terminology, \citet{makri_coming_2012} studied experienced serendipity, whereas \citet{bjorneborn_three_2017} described afforded serendipity. 
{In our study, we will focus exclusively on experienced serendipity since the other levels are contingent on it; without understanding what experienced serendipity entails, we cannot study what affords it, nor its strategic benefits.}

Furthermore, \citeauthor{smets_serendipity_2022-1} argues that ``there is in fact no single definition, nor a unique operationalization of serendipity; both are dependent on the specific context''~\cite{smets_serendipity_2022-1}.
{In other words, a user's experienced serendipity with RecSys may be different from their experienced serendipity in a library or when making a scientific discovery.}

\subsection{Serendipity in RecSys}

Researchers in RecSys have been interested in serendipity since the field's inception.
In recent years, this interest has grown significantly~\cite{ziarani_serendipity_2021}, as serendipity became a central focus of the increasingly popular beyond-accuracy paradigm~\cite{mcnee_being_2006}.
The concept of serendipity in RecSys is linked to several potential benefits such as increasing user satisfaction~\cite{asikin_stories_2014,maccatrozzo_everybody_2017}, promoting horizon broadening, counteracting filter bubbles~\cite{wang_impacts_2020}, and supporting self-actualization by helping users explore and develop their tastes~\cite{graus_let_2018,knijnenburg_recommender_2016,sullivan_reading_2019}.

Despite the significant attention paid to serendipity in RecSys, we argue that users’ experiences of serendipity with RecSys are still poorly understood.
This is evidenced by inconsistent definitions and operationalizations found in the literature~\cite{binst_how_2024,kotkov_rethinking_2023,ziarani_serendipity_2021};  
{several notable authors in the field mention this conceptual ambiguity surrounding serendipity in RecSys as a major obstacle for studying it~\cite{ziarani_serendipity_2021, kotkov_challenges_2016, kotkov_dark_2024}.}
While most studies emphasize \textit{unexpectedness} or \textit{surprise} as a core component~\cite{ziarani_serendipity_2021}, other studies highlight additional components, including \textit{novelty}~\cite{wieland_one_2021}, \textit{difficulty-to-discover}~\cite{de_pessemier_comparison_2014}, \textit{unusualness}~\cite{stitini_towards_2023} and \textit{resonance}~\cite{yamaba2013serendipity}.
Furthermore, some popular definitions contain inherent contradictions: Serendipity is often defined as encountering items that differ from a user’s profile~\cite{kaya_novel_2022,yu2018accuracy}, yet it is also described as encountering relevant items, which are typically defined as similar to a user’s profile~\cite{kotkov_survey_2016,pastukhov_offline_2022}.
This duality suggests that serendipity must simultaneously involve both irrelevant and relevant items{~\cite{ziarani_serendipity_2021,kotkov_survey_2016}}.

Several works have sought to address this conceptual ambiguity~\cite{kotkov_survey_2016, kotkov_rethinking_2023, ziarani_serendipity_2021}. 
Both \citet{ziarani_serendipity_2021} and \citet{kotkov_survey_2016} conducted a literature review focusing on serendipity in RecSys and identified its most common components in the literature. 
Both found that serendipity in RecSys is most often conceptualized as the intersection of \textit{relevance}, \textit{unexpectedness} and \textit{novelty}. 

However, even papers that agree on this \textit{conceptualization} of serendipity adopt different \textit{operationalizations} of these components relating to very different {expressions} of serendipity{~\cite{kotkov_dark_2024,ziarani_serendipity_2021}}. 
{For example the most agreed-upon component---unexpectedness---has been operationalized as the dissimilarity between a recommendation list and what a primitive RecSys recommends~\cite{murakami_metrics_2008,ge_beyond_2010}, the dissimilarity between a recommendation list and the user profile~\cite{kaminskas2016diversity}, or as the inverse of popularity~\cite{karpus2017serendipitous}.
Alternatively, it has also been measured through user-reported surprise or the perception of recommendations as unexpected~\cite{jenders2015serendipity,chen2019serendipity}.}


{This example illustrates what we label as a conceptualization-operationalization gap.}
Although serendipity is most often conceptualized as a subjective user experience, the overwhelming majority of papers surveyed by \citeauthor{ziarani_serendipity_2021} and \citeauthor{kotkov_survey_2016} evaluate serendipity in RecSys using objective measures of system performance.
Thus, {we argue that} they provide a biased and limited view of serendipity in RecSys focused on \textit{afforded serendipity}, not \textit{experienced serendipity}, the subject of this work. 


{In response to the narrow definition of serendipity in RecSys as the intersection of relevance, unexpectedness and novelty, \citet{kotkov_rethinking_2023} propose a broader definition, termed \textit{generalized serendipity}}.
For this definition, \citeauthor{kotkov_rethinking_2023} draw from theoretical work by~\citet{yaqub2018serendipity} that conceptualizes serendipity in the context of how scientific discovery is made, distinguishing between Mertonian, Walpolian, Stephanian and Bushian serendipity.
Each of the four types of serendipity assumes that serendipitous discoveries help researchers reach goals in an unplanned manner.
\citet{kotkov_rethinking_2023} adapts this conceptualization to RecSys, defining generalized serendipity as: \textit{``an item is serendipitous if it helps the user to achieve at least one goal different from those they set out to achieve with the recommender system.''}


In their follow-up work, \citet{kotkov_dark_2024} explore how the conceptualization based on relevance, unexpectedness and novelty---which they terms \textit{Recsys Serendipity}---differs from serendipity as reported by users when asked if they found an item serendipitous---termed 
\textit{User Serendipity}---and their new definition of \textit{Generalized Serendipity}.
They find that there are considerable differences between items experienced as RecSys, User, or Generalized Serendipity, exposing what they call the \textit{dark matter} of serendipity. 

{To explore this \textit{dark matter} of serendipity, we argue that we should study \textit{What constitutes experienced serendipity from the user's perspective? What are its main components?} and \textit{What conditions underlie them?}
To the best of our knowledge, this is the first study to address this gap by conceptualizing users' experienced serendipity in the context of RecSys.}



\section{Methodology} \label{sec:meth}
To understand {how users experience serendipity} {with RecSys}, we conducted semi-structured interviews with 17 participants, in Belgium {between July and December 2024}.
{Subsequently, we} analyzed the data obtained from these interviews using the grounded theory framework~\cite{strauss_basics_2015}. 
Below, we describe our sampling approach (Section \ref{subsec:meth-sample}), interview protocol (Section \ref{subsec:meth-prot}), data processing, and analysis (Section \ref{subsec:meth-analysis}).  

\subsection{Sampling Approach} \label{subsec:meth-sample}



{Our conceptual framework aims to capture experienced serendipity with RecSys among active users aged 20–60. 
To ensure diversity in our sample of participants, we applied sampling quota by age and gender~\cite{campbell2020purposive}.
Quota sampling is a form of purposive sampling, which is a standard practice in qualitative research and grounded theory~\cite{campbell2020purposive,strauss_basics_2015,guest_how_2006}.
Further, we selected participants who self-report to frequently engage with platforms that utilize RecSys.
All participants, as well as the interviewer, reside in Belgium.
Despite this, we expect our findings to generalize beyond Belgium, as participants primarily discussed large international platforms.
Finally, we note that some of the participants were familiar with the interviewer.
Despite this, we argue that this familiarity does not bias our results, as these participants were unaware of the interviewer's research focus on serendipity and thus had no knowledge of their ideas on the subject.
Table~\ref{tab:sample-description} provides an overview of the participants, their self-reported characteristics, and the platforms they discussed in their interviews.}
{We continued recruiting participants until we reached theoretical saturation, i.e., no new codes or relationships between codes emerged from subsequent interviews~\cite{guest_how_2006,strauss_basics_2015}.}

\begin{table*}

  \centering
  \begin{tabular}{ccc p{10cm}}
    \toprule
    Participant & Age Category & Gender & Platforms Discussed \\
    \midrule
    P1 & 20--30 & M & Amazon Prime, Bol.com, Spotify, YouTube, TikTok, Instagram \\
    P2 & 20--30 & M & Spotify \\
    P3 & 31--40 & F & Amazon, Netflix, Facebook, Albert Heijn App, Spotify, Instagram \\
    P4 & 20--30 & M & Zalando, Spotify, Instagram, Amazon \\
    P5 & 20--30 & M & YouTube \\
    P6 & 20--30 & M & Instagram, YouTube \\
    P7 & 20--30 & F & Netflix, Spotify, TikTok \\
    P8 & 20--30 & M & De Bib, Instagram, Reddit, TikTok, YouTube \\
    P9 & 20--30 & M & Netflix, Spotify, Pinterest, TikTok, Zalando \\
    P10 & 20--30 & M & Facebook, YouTube, Booking.com \\
    P11 & 31--40 & M & Zalando, YouTube, Spotify, Facebook, Instagram \\
    P12 & 20--30 & F & Spotify, Instagram, E-Commerce, Facebook \\
    P13 & 51--60 & F & Facebook, Spotify, Netflix \\
    P14 & 41--50 & F & Pinterest, Instagram, Netflix \\
    P15 & 51--60 & M & LinkedIn, Instagram, Netflix \\
    P16 & 41--50 & M & Spotify, Twitter, Pinterest, Instagram \\
    P17 & 41--50 & F & Spotify, Instagram, Pinterest \\
    \bottomrule
  \end{tabular}
    \caption{Overview of the personal characteristics of and platforms discussed by the participants in our study.\label{tab:sample-description}}
\end{table*}

\subsection{Interview Protocol} \label{subsec:meth-prot}

\paragraph{Consent} \label{para:consent}
{Following our institute's data protection guidelines, participants received an informed consent document outlining the study’s purpose, data handling practices, and privacy safeguards.}

\paragraph{Interview Approach}
Because many participants lacked preconceived notions about serendipity, the concept had to be collaboratively constructed. 
To facilitate this, we employed the active interview approach~\cite{holstein1995active}. 
Unlike traditional methods where participants passively respond to questions, this approach encourages interviewees to actively steer the conversation, while interviewers stimulate interpretation and meaning-making~\cite{holstein1995active}.

\paragraph{Interview Questions}
Establishing an understanding of experienced serendipity was complicated due to participants' limited understanding of what RecSys are.
Therefore, we began our interviews with questions that assess their `RecSys literacy' and provided explanations when necessary.
{We then explored their interactions with RecSys, asking about the specific systems they regularly used.}

Finally, the discussion shifted to the study's core topic: experienced serendipity. 
To explore this topic, participants were first asked about {specific} interesting {or valuable} discoveries they made while using {the previously mentioned} RecSys. 
Follow-up questions, such as \textit{Why was it interesting?} \textit{What did you like about it?} and \textit{How did you discover it?} delved deeper into the context and significance of these discoveries{, and aimed to explore how experienced serendipity expresses itself}.
Additionally, we also inquired about negative experiences with recommendations to identify components and conditions that hinder experiencing serendipity. 

\paragraph{Interview Process}
The interviews ranged from 13 to 78 minutes in duration. 
Early sessions were exploratory, while later ones built on previous findings, resulting in greater depth and length. 
{All participants were interviewed in Dutch. 
Quotes used in this paper were translated to English by the authors.}

\paragraph{Interview Transcripts}
{To ensure transparency, the dataset underlying our conceptual framework will be made available upon request via Zenodo\footnote{\href{https://doi.org/10.5281/zenodo.15131547}{doi:10.5281/zenodo.15131547}}.}


\subsection{Grounded Theory} \label{subsec:meth-analysis}
{To develop our conceptual framework, w}e employed grounded theory, a widely recognized inductive analysis method designed to understand phenomena by grounding them in qualitative data~\cite{strauss_basics_2015}.
{Grounded theory} is considered well-suited for generating original insights~\cite{mehmetoglu2006examination} and providing a rich description of concepts~\cite{matteucci2017elaborating,woodside2004grounded}, making it an ideal choice for exploring experienced serendipity.

{Grounded theory analysis involves multiple iterations of interviewing, analyzing, and refining a theoretical framework \cite{strauss_basics_2015,muller2012grounded}.} 
Each iteration tries to address weaknesses in the theory developed in the previous iteration---a process called `abduction'\cite{muller2012grounded}. 
This process continues until an iteration produces a theory that remains unchanged by new data, i.e., no new codes emerge that alter or expand the core concepts of the framework, indicating that theoretical saturation has been reached \cite{muller2012grounded,strauss_basics_2015}. 
In our case, theoretical saturation was reached during the final iteration of the grounded theory process; the last three interviews did not alter the conceptual framework. To validate this, we presented the framework to participants at the end of the final two interviews. Although these interviews yielded valuable examples of experienced serendipity, they led to no revisions, confirming that theoretical saturation had been achieved~\cite{guest_how_2006}.

{The grounded theory} analysis process involves two main phases: open coding and axial coding. 
In the open coding stage, we conceptualized serendipitous experiences with RecSys by identifying their properties,  e.g., \textit{enriching}, and dimensions in the data, i.e., elements that add variation, such as \textit{resonance}.
In the axial coding stage, we linked concepts together by clustering conditions into main components. For example, \textit{unusual}, \textit{novel} and \textit{taste reincarnation} were clustered together under \textit{refreshing}.
{For this coding process, as well as documenting our thought process, we used MAXQDA24.}

\section{Results} \label{sec:results}

In this Section, we first describe the conceptual framework that arose from our analysis (Section \ref{subsec:framework}). 
Subequently, we demonstrate how our framework aids to classify and describe experienced serendipity by means of positive and negative experiences and edge cases discussed by our participants (Section \ref{subsec:cases}). 
{We note that our participants discussed a variety of different platforms originating from many different application domains of RecSys. 
{However, we also note that the majority of platforms discussed were large, international platforms. 
Very few participants gave examples of experienced serendipity on small, local platforms.}}

\subsection{A Conceptual Framework of Experienced Serendipity} \label{subsec:framework}
Based on our grounded theory analysis, we conceptualize experienced serendipity as 
\textit{a user experience in which the user unintentionally encounters content that feels fortuitous, refreshing, and enriching}.
Together, these three criteria are necessary and sufficient for classifying a user's experience as serendipitous.
We therefore term them the \textbf{main components} of experienced serendipity.
{These main components can be satisfied through different \textit{conditions}.}
{Below we give a summary of the resulting framework:}
\begin{itemize}
    \item \textbf{Fortuitousness:} The user's experience must involve an \textit{unintentional} encounter with content that is either \textit{difficult-to-find}, \textit{unexpected}, or \textit{uncovers a subconscious interest}.
    \item \textbf{Refreshing:} The user's experience must involve encountering content that is either \textit{novel}, \textit{unusual}, or reignites a past interest through \textit{taste reincarnation}. 
    \item \textbf{Enriching:} The user's experience must be \textit{intriguing}, \textit{inspiring}, \textit{impactful}, \textit{relevant}, or \textit{resonating}, while none can be negative. 
\end{itemize}

This type of conceptualization---where not all conditions must be present---is known as family resemblance conceptualization~\cite{goertz_social_2006,wittgenstein_philosophical_2009}.
It accommodates the slippery nature of serendipity, enabling us to differentiate between different flavors of experienced serendipity through various combinations of conditions. 
In what follows, we discuss each of the main components and their different conditions in detail, {drawing upon quotes from our interviews.}
We selected the quotes that best illustrate how the developed theory is grounded in users' experiences, as is customary in grounded theory \cite{strauss_basics_2015}.

\begin{table*} 
    \begin{tabular}{l|l|l|l}
        \toprule
        \textbf{Main Component} & \textbf{Condition} & \textbf{Expression} & \textbf{Opposite} \\
        \midrule
        \multirow{5}{2.5cm}{\textit{\textbf{Fortuitous vs. \newline Anticipated}}} & 
        \textit{Unintentional} & Encountered & Explicitly searched \\
        & Difficult-to-find & Recsys facilitated discovery & Easily discovered on one's own \\ 
        & Uncovering & Revealing latent interests & Related to conscious interests \\
        & Unexpected & Random & Predictable \\
        \midrule
        \multirow{4}{2.5cm}{\textit{\textbf{Refreshing vs. \newline Boring}}} & 
        Novel & Completely unknown & Consumed\\
        & Unusual & Different compared to regular consumption & Hyperfixation \\
        & Taste reincarnation & Reconnecting & Weakening connection \\
        \midrule
        \multirow{7}{2.5cm}{\textit{\textbf{Enriching vs. \newline Brainrot}}} & 
        Intriguing & Sparks curiosity & Uninteresting \\
        & Inspiring & Motivating & Demotivating \\
        & & Sparks creativity & Stifles creativity \\
        & Impact & Positive influence on life & Negative influence\\
        & Relevance & Empowering & Disempowering \\
        & & Useful & Useless/Harmful\\
        & Resonance & Touching & Out-of-touch \\
        \bottomrule
    \end{tabular}
    \caption[Conceptualization of experienced serendipity]{{Schematic overview of our conceptualization of experienced serendipity. experienced serendipity has three main components---fortuitous, refreshing and enriching---and different conditions belonging to each of the main dimensions. We use \textit{italics} to indicate necessary components of experienced serendipity.}  \label{tab:serendipity}}
\end{table*}

\subsubsection{Fortuitous vs. Anticipated}
The \textbf{fortuitous} component captures the role of chance in content discovery. 
It is conceptualized as the unintentional discovery of content, a necessary condition, that is either difficult-to-find, unexpected, or uncovers a latent information need. 
The opposite of fortuitous is \textbf{anticipated}, which occurs when content is explicitly searched for, or when it is easy-to-find, expected, and meets a conscious information need. 

\paragraph{Unintentional:} 
refers to encountering content in an unplanned way, as opposed to actively searching for it. 
For example, a user might come across a new song via YouTube's autoplay or Spotify's song radio.
{The absence of intentionality when encountering something is universally considered to be a necessary condition for serendipity, e.g., ~\cite{makri_coming_2012, smets_serendipity_2022-1, smets_serendipity_2022, bjorneborn_three_2017}.}
Hence, since we start from this assumption when interviewing our participants, we do not include a specific quote to illustrate this here, as this condition is exemplified through all the following quotes.

\paragraph{Difficult-to-find:} 
refers to content that users perceive as difficult to discover independently. 
RecSys can assist by navigating the item space, thereby facilitating discovery. 
Take for example, when \textbf{P3} was looking for a clock radio on Amazon:  \textit{``if you don't know exactly what you want, these recommendations are quite handy; you search a bit; it gives you recommendations, and in that way, you get to discover more interesting items.''}

\paragraph{Uncovering:} 
involves finding something you did not know you were seeking. 
Often, people have subconscious information needs and latent interests, which RecSys can help uncover.
For instance, \textbf{P11}: 
\textit{``The last two months I had this itch. The music I listened to doesn't do it anymore; I am getting bored of it. Should I listen to some Psytrance? No, that’s not it. Should I listen to downtempo? No, also not it. Searching, searching, searching until I encounter this post of Behemoth on Instagram: `Ah I haven’t listened to black metal for a while!' and indeed, that was it, my itch.''}

\paragraph{Unexpected:} 
content refers to encounters that feel random or coincidental. 
This happens when users do not understand why content was recommended. 
For instance, \textbf{P8} recalls: \textit{``recently I stumbled across this super random geography subreddit---something I’m not involved with at all---yet I clicked on it, and it was interesting.''} 


\subsubsection{Refreshing vs. Boring}
This component captures how \textbf{refreshing} the encountered content feels. 
Content is considered refreshing when it is novel, unusual, or triggers taste reincarnation.
Conversely, the opposite of a refreshing user experience is a \textbf{boring} one. 
This occurs when most of the encountered content is already familiar, closely resembles what you regularly consume, and fails to spark taste reincarnation.

\paragraph{Novel:} 
refers to content ranging from completely unknown, to familiar but not yet consumed, and finally to content that has already been consumed. 
Novelty does not require the content to differ from a user’s profile. 
In fact, serendipity often arises when users discover novel content that aligns closely with their preferences. 
For example, when asked about series that stood out to her, \textbf{P13} stated: \textit{``Actually, they are series I hadn’t heard about before, but that are in the same genre I like to watch.''}

\paragraph{Unusual:}
content differs from what a user typically consumes, for example, in genre, style, or perspective. 
\textbf{P8} mentions this need for unusual content, and vents his frustration about when it is not met:
\textit{``it [the algorithm] hyperfixates immediately once you like a couple of reels of the same topic. The algorithm knows then: ‘okay, he likes this. Bam! I’ll spam the heck out of him with that. That is really annoying. Sometimes you want something new, something unusual, because you feel trapped.'''}
This sense of restriction is echoed by \textbf{P4}, who compares it to:
\textit{``as if I'm wearing blinkers; I get trapped in an isolated part of the item catalog.''}

\paragraph{Taste reincarnation:} 
involves encountering content that reignites interest in something the user has not engaged with in a while, {like \textbf{P2} explains: \textit{``With Discover Weekly, you’re sure that you won’t hear a song you’ve already played on Spotify, but you know much more music than what you’ve listened to on Spotify. So if there’s a song in my Discover Weekly that I actually know, but I haven’t heard it in years, then I think that's pretty cool.''}
\textbf{P15} who studied history, gives a concrete example of taste reincarnation}:
\textit{``I have some former classmates in my network [on LinkedIn], and some of them are still into history. Occasionally, I encounter one of their posts and think, `Ah that is intriguing!', inspiring me to dig deeper into the subject again.''}.
Another example was provided by \textbf{P8}, who rediscovered Taylor Swift through a TikTok trend featuring the song 'Lovestory':
\textit{``I remembered I liked that song, so I decided to look her up again. That’s how I started listening to her again."}
Motivating him to collect her LPs, immerse himself in her lyrics, and travel to one of her concerts.

\subsubsection{Enriching vs. Brainrot}
This component reflects how meaningful and valuable the encountered content feels to the user. 
Encountered content is \textbf{enriching} when it is perceived as either intriguing, inspiring, impactful, resonating or relevant by the user. 
The opposite of enriching content is \textbf{brainrot}, which is experienced when users encounter content that is either uninteresting, has a negative influence on your life, captures you in a boredom bubble, is disempowering, harmful, or out-of-touch.
An important difference from the other two main components is that a user's experience is only enriching if none of the opposite expressions are present.
For example, if encountered content is harmful and motivating at the same time---as may be the case for some posts encountered on social media---it cannot be considered enriching.

\paragraph{Intriguing:} 
content sparks curiosity.
Users are intrigued by content if they start delving deeper into it, like \textbf{P8}: \textit{``When I find an interesting post on Reddit, I click on it; I follow the hyperlinks to, for example, Wikipedia; I read that Wikipedia page; and that's how you end up learning more about topics.''}
Interestingly, we observed an interaction between the conditions unusual and intriguing; unusual recommendations are appreciated only if the content is also intriguing. 
In the words of \textbf{P13}:
\textit{``Like Squid game: It is totally different from what I normally watch but I really enjoyed it. It [serendipitous content] can be something unusual, but it must be intriguing.''} 

\paragraph{Inspiring:} 
content can express itself as a spark of creativity. 
For example, \textbf{P14}, an interior architect, came across an image of a kitchen with distinctive square tiles while browsing Instagram:
\textit{``There are other images of similar tiles, but this is a natural stone---and it’s not so much the natural stone that makes it beautiful---but the way it is cut. So that was truly an inspiring discovery for me.''}

Inspiration can also manifest itself as motivation. 
This occurs when the content encountered energizes users to achieve goals, like pursuing fitness milestones or exploring new hobbies. \textbf{P6} illustrates this:
\textit{``That [bouldering] is something I discovered and that I really want to try now.''} 

Contrastingly, interacting with RecSys can also foster laziness and stifle creativity. 
This happens when users get caught in a \textit{boredom bubble}~\cite{michiels2024methodologies}.
\textbf{P8} expresses this sentiment vividly:
\textit{``After a while you just think ‘What the F--? What am I doing with my life?’ and you feel very bad for a moment, lying on your bed or on your sofa, feeling like a couch potato.''}

\paragraph{Impact} Positive impact occurs when users encounter content that improves their life. 
This impact can range from life-changing events, such as discovering a dream house on Facebook, to smaller discoveries, as \textbf{P14} explains:
\textit{``Something really silly: how do I cut a pineapple? I used to cut it differently until I accidentally came across a video on social media showing me a new way.''}

However, many respondents struggle to recall recommendations that had a positive impact, often stating that most suggestions failed to leave a meaningful influence. 
In contrast, some users highlighted the negative impact of certain platforms, pointing to their addictive nature and potential harm to self-esteem. 
\textbf{P12} illustrates this sentiment:
\textit{``Toxicity and social media almost go hand in hand. Especially now that I am in this self-help content bubble, it's easy to think: `Oh, I don’t look like all those yoga chicks; maybe I need supplements too.' ''}

\paragraph{Relevance:} 
refers to encountering content that directly addresses a user's information needs. 
This can manifest itself as an empowering experience that increases self-efficacy. 
For example, \textbf{P14} explains how finding inspiring interior images on social media supports her professional work:
\textit{``You can show others what you have in mind for them,''} 
which facilitates collaboration with clients and empowers her in her role. 
Conversely, recommendations can also be disempowering as we will illustrate {in Section \ref{subsec:cases}}. 

Relevance can also manifest {itself} as encountering content that is perceived useful.
\textbf{P17} shares an example:
\textit{``I suddenly encountered this Ikea Hack: ‘Oh, that is useful [for my renovation]’. So, I saved it for later.''} 
{Conversely, recommended content may be perceived as harmful, such as when users encounter misinformation that leads to poor decision-making.
\textbf{P16} expresses this sentiment about posts encountered on X: \textit{``the content there is really trash; those recommendations are driven by sensationalism.''}}

\paragraph{Resonance:} 
entails encountering content that touches you, that you connect with. 
For example, \textbf{P10} mentions:
\textit{``There are times when I’m sad, and I go on Facebook, and there’s this perfectly timed quote for when you’re feeling low---and it really resonates with me; thank you Facebook for understanding me.''}

The reverse of resonance occurs when recommendations feel out-of-touch, leaving users with the impression that the RecSys fails to understand their tastes. 
For example, \textbf{P3} describes using the song radio feature on Spotify after watching a movie:
\textit{``It sometimes makes these awkward associations; it doesn't consider that the reason I like that song isn't because it's connected to a movie''}

\subsection{Different Flavors of Experienced Serendipity} \label{subsec:cases}
{In this Section, we demonstrate how our conceptualization is capable of classifying the various flavors of experienced serendipity, as well as user experiences {that lack serendipity or} even directly oppose it.} 
This discussion is not exhaustive, but illustrates how our conceptualization, as shown in Table \ref{tab:serendipity}, provides the building blocks and rules to accurately capture experienced serendipity.
{The conditions satisified by the flavor of serendipity are shown in \textbf{bold}. What we consider to be the `core characteristic' of the flavor is \underline{underlined}.}
{This core characteristic exemplifies this flavor of serendipity, while other conditions may vary.}

\subsubsection{Positive Experiences}

{In this Section, we present specific cases of different flavors of experienced serendipity.
Following our conceptual framework, experiencing serendipity is by definition a positive experience that entails fortuitously encountering content that is both refreshing and enriching.}

\paragraph{Life changing serendipity}
\textbf{P3} mentioned discovering a documentary about twins and the nature versus nurture debate, which she encountered \textbf{unintentionally} on the top of her Netflix home screen:
\textit{``That happens quite often; Netflix does show things at the top that are atypical. But sometimes, those things are really bad. Still, they are always something different.''}

This discovery was \textbf{refreshing} because it was \textbf{unusual} and \textbf{novel}:
She had not heard of this documentary before and typically watched fantasy series on Netflix. 
She was intrigued by the documentary because it focused on twins, a subject that fascinated her since studying psychology.
Most importantly however, it was {\textbf{enriching} because it was \underline{{\textbf{impactful}}} and \textbf{inspiring}}: {Observing that a twin with a vegan diet was significantly healthier than their carnivorous counterpart inspired her to adopt a vegetarian lifestyle.}

\paragraph{Relevant but difficult-to-find serendipity}
\textbf{P14} shared a serendipitous experience where she stumbled upon a design solution to a problem she was facing:
\textit{``I was looking for a way to open up a garden room, and I eventually came across this [shows picture]. I think it’s really cool because I was looking for something like this, but I wouldn’t have found it on my own, so that’s really interesting.''}

She discovered the idea while browsing Pinterest, which she frequently uses for inspiration. 
{The encounter was \textbf{fortuitous} because the content was \underline{\textbf{difficult-to-find}}---it is unlikely she would have discovered this solution without the platform's help---and \textbf{unintentional}.}
The \textbf{novelty} of the idea made the discovery \textbf{refreshing}, while its \underline{\textbf{relevance}} to her project made it \textbf{enriching}. The solution was not only \textbf{useful} for her current project but also \textbf{empowered} her professionally:
\textit{``Even if I had come up with this idea myself, I wouldn’t have been able to show it to my client and say, ‘This is what we want to build.'''} 

\paragraph{Taste-broadening serendipity} 
\textbf{P4} experienced a moment of taste-broadening serendipity when he encountered a shoe recommendation on Zalando: 
\textit{``It’s strange because I never specifically looked for those shoes, but I ended up buying them and they pleasantly surprised me because it was a good deal and they are really beautiful.''} 

{This recommendation is experienced as \textbf{fortuitous} because it was \underline{\textbf{unexpected}}: To \textbf{P4}, the encounter felt strange and somewhat random.}
The recommendation was also \textbf{refreshing}, as it introduced him to a style he wouldn’t normally consider:
\textit{“They were these moccasin-type shoes that I would never have bought on my own, but I’m really happy that I did.”}
{In other words, the recommendation was \underline{\textbf{unusual}} for \textbf{P4}.}
{At the same time, it \textbf{enriched} his experience—the shoes \underline{\textbf{intrigued}} him because they were on sale, prompting him to buy them.}
{After purchasing them, their style gradually \underline{\textbf{resonated}} with him, leading to a taste-broadening experience.}

\paragraph{Taste-deepening serendipity}
Experiencing serendipity does not always broaden taste;
instead, it can deepen an existing appreciation.
\textbf{P11} provides an example: 
\textit{``Oh, yes song radio on Spotify is absolutely fantastic, much better than Discover Weekly. Song radio has helped me discover so many bands. For example, I discovered this amazing song by a new wave band from Belarus. I thought, `Oh wow, this is great. Let me check out the song radio.' And through that song radio, my friends and I really got into new wave, and now we even go to new wave parties. That’s also how we discovered a festival in The Hague, which we now attend every year.''} 

{This encounter is classified as \textbf{fortuitous} because the discovery was \textbf{unintentional} and \textbf{difficult-to-find}---\textbf{P11} believes he would not have discovered these bands on his own.}
{It is also \textbf{refreshing} because the recommendation was \textbf{novel}; it introduced him to a \textbf{previously unknown} band.
However, what defines this flavor of experienced serendipity is that the genre itself is \underline{\textbf{not unusual}}; it aligns with \textbf{P11}'s existing music taste.}
{However, it is deeply \textbf{enriching}, as the content strongly \underline{\textbf{resonated}} with him, \textbf{inspired} him, and \textbf{motivated} him to attend the festival in The Hague.}


\subsubsection{Edge Cases}

{In this Section, we present some `edge cases' that lack one or more core components of experienced serendipity. 
However, unlike negative experiences, they do not score negatively on the enriching conditions. 
Although these instances can provide a pleasant user experience, they do not evoke the magic of experiencing serendipity.}

\paragraph{Predictable content}
{\textbf{P15} explains that most of the content he discovers on LinkedIn {may be interesting, but} is not really fortuitous because it is \textbf{\underline{predictable}} and \textbf{not difficult-to-find, nor uncovering}.:}
{\textit{``And on LinkedIn... Pff, I don’t really get surprises there either. What I get recommended, I kind of expect---topics from my field, like social security, labor inspection, EU stuff. It’s not stuff I search for, but it’s not unexpected either. I can’t really think of a moment on LinkedIn where I thought, 'Wow, I didn’t expect this, and I actually did something with it.'''}
} Although the recommendations \textbf{P15} encountered were \textbf{somewhat enriching}, as they were \textbf{interesting}, and \textbf{refreshing} due to their \textbf{novelty}, he remains unconvinced of their overall value.:
\textit{``It’s just interesting, so I read it. And yes, it’s relevant. But it’s limited. I wouldn’t say it’s something I’d miss if I hadn’t seen it.''}

\paragraph{Difficult-to-find vs. Hyped}

{\textbf{P9} aptly illustrates the difference between a `good' and `serendipitous' recommendation:}
{\textit{``Avatar was already hyped before, so I definitely wanted to see that. I had heard a lot about it. But One Piece I didn’t know. I never read the manga or anything. I had once heard about it in the cinema, how people liked the mangas, but that was three or four years ago. And then they made a live-action series. It looked really good in the trailer, and I thought, “Oh sh**,” the effects and everything were great.''}
}

Although \textbf{P9} encountered both Avatar and One Piece \textbf{unintentionally}, and both were \textbf{enriching} and \textbf{refreshing}---as they \textbf{resonated} with his taste and \textbf{were new to him}---Avatar did not evoke experienced serendipity.
He saw it as a \underline{\textbf{predictable}} recommendation, \textbf{easily discoverabl}e due to the surrounding hype.
In contrast, One Piece felt \textbf{unexpected} and was something he would have had \underline{\textbf{difficulty finding on his own}}.

\subsubsection{Negative Experiences}
{This paragraph presents negative experiences with recommender systems and relates them to our conceptual framework. 
These cases show how negative experiences can be captured using the opposites of the enriching conditions.}

\paragraph{Hyperfixation} 
\textbf{P6} described an instance where a RecSys fixated on a topic that {no longer} resonated with the user. This resulted in a restrictive and disempowering experience: 
\textit{``Recently, I’ve liked a few posts because I was going through a bit of a sad period. After a day or two, my Explore page was filled with depressing reels and posts. I’m fed up; I don’t want to see this stuff anymore. Suggest something else to me---make me think about something different---show me fun things, interesting things.''}

The algorithm's \underline{\textbf{hyperfixation}} resulted in highly \textbf{predictable} content that failed to address his evolving needs. 
Instead of offering new or unusual content, the system restricted his experience of the item space by repeatedly recommending the same type of depressing, \textbf{irrelevant} posts, which felt \textbf{out-of-touch}, \textbf{harmful}, and \textbf{disempowering} by further amplifying his negative feelings, driving him to quit the platform.

\paragraph{Brainrot} 
\textbf{P16} regularly scrolled through Twitter, but the platform's transformation into X left him disappointed:
\textit{``The content there is really trash; those recommendations are driven by sensationalism. Meanwhile, I’m trying to curate my profile around a different type of content, but it just gets flooded with junk.''}
 
According to \textbf{P16}, the platform’s algorithms emphasize sensationalist and polarizing content, offering a predictable and restrictive worldview. 
Rather than discovering new and enriching material, he encountered content that felt \textbf{irrelevant}, \textbf{out-of-touch}, and \textbf{uninteresting}. 
The \underline{\textbf{lack of enriching content}} and control over his feed ultimately led him to leave the platform.

\paragraph{Disempowering recommendations} 
\textbf{P17}, who enjoys knitting as a hobby, occasionally encounters knitting content on Instagram. 
However, her experience is not always positive: 
\textit{``sometimes I think, oh no, that’s too hard, it’s not going to work. Like, how do people even have time for that, and then you start to feel disempowered.''} 
The content she encountered on Instagram set an unrealistically high standard, leaving her feeling \underline{\textbf{disempowered}}, \textbf{demotivated}, and \textbf{less connected} to knitting.

\section{Discussion} \label{sec:discussion}

Serendipity remains elusive in RecSys research, plagued by conceptual ambiguity leading to inconsistent---even contradictory---operationalizations across the literature.
While several works have sought to address these issues, none have focused specifically on how users experience serendipity with RecSys. 

In this paper, we addressed this research gap and investigated the question---\textit{How do users experience serendipity with RecSys?}---through in-depth interviews with frequent users of online RecSys.
Using a grounded theory approach, we developed a conceptual framework for experienced serendipity in the context of RecSys. 

\paragraph{Serendipity in RecSys is often defined too narrowly}
Our findings confirm that many elements identified in previous literature are relevant to define experienced serendipity, e.g., novelty, relevance and unexpectedness---the components of `RecSys Serendipity”.
However, our findings also show that not all of these components are necessary for experiencing serendipity;
while experienced serendipity is always fortuitous, refreshing, and enriching, these main components can be expressed through different combinations of conditions, of which the combination of novelty, unexpectness and relevance is only one.

\paragraph{Flavors of experienced serendipity in RecSys}
Instead, these different combinations of conditions produce \textit{different flavors of serendipity}. 
In Section \ref{subsec:cases}, we highlighted some flavors that have remained underexplored due to the overly narrow conceptualization of serendipity in the literature.
For example, a user may perceive an experience as refreshing when they encounter a new song within their preferred genre that resonates deeply with their mood---a phenomenon we call \textit{taste-deepening serendipity}.
Alternatively, a user may feel refreshed by discovering an entirely new genre that intrigues them---which we call \textit{taste-broadening serendipity}.

\paragraph{RecSys serendipity lacks depth}
Not only did we find the definition of RecSys serendipity too narrow, at the same time, it lacked depth, leading to inconsistent operationalizations in the literature.  
For example, the many operationalizations of \textit{unexpectedness} can largely be explained as belonging to different conditions of the main components \textit{fortuitous} and \textit{refreshing}: The inverse of popularity likely leads to content that is \textit{difficult-to-find}, whereas different from the user profile can be considered \textit{unexpected}, as it feels random---instead of predictable---to the user, but mostly \textit{unusual}, as it is different from what the user regularly consumes.  
Thus, beyond broadening the conceptualization of experienced serendipity, our framework also adds depth to the concept, exploring what \citeauthor{kotkov_dark_2024} call \textit{the dark matter} of serendipity, by identifying these distinct conditions that richly describe and make experienced serendipity more tangible~\cite{gerring1999makes}.
Our in-depth conceptualization clarifies the concept and resolves many of the apparent contradictions between and within definitions of serendipity.
For example, our framework shows that the contradiction between relevance---similar to the user’s profile, and unexpectedness---different from the user’s profile, is a result of conflating two different flavors of serendipity,  \textit{taste-deepening} and \textit{taste-broadening} serendipity.

\paragraph{Conceptual framework enables consistent operationalization}
Furthermore, our conceptual framework establishes a foundation for operationalizing experienced serendipity consistently and correctly. 
While current research frequently operationalizes serendipity by either summing its main components, for example~\cite{asikin_stories_2014,pastukhov_offline_2022,pardos2020designing}, or analyzing them separately~\cite{stitini_towards_2023,saito2022fusion}---these approaches are problematic based on our conceptualization.
Since the main components are necessary, they must always be present for experienced serendipity to occur.
A formula that calculates experienced serendipity by summing the components fails to ensure this:
one component can score exceptionally high, masking a low score in another~\cite{goertz_social_2006}.
Similarly, analyzing the components separately overlooks their interdependence.
One possible way to avoid this problem is by setting a minimum threshold score for each main component.

\paragraph{Enables designing RecSys that afford serendipity}
Beyond theoretical contributions, our framework also offers practical insights for RecSys developers seeking to design RecSys that afford serendipity.
In particular, the conditions we identify clarify which perceptions and experiences should be triggered to foster experienced serendipity.
For example, the condition of \textit{taste reincarnation} may inspire RecSys to recommend items that users are familiar with but have not engaged with for a long time.
Similarly, unusual recommendations should be made intriguing---such as by highlighting them---to increase the potential for experiencing serendipity.

\section{Conclusion}
In this study, we conducted a user-centered evaluation of experienced serendipity through semi-structured interviews with {a diverse\footnote{diverse in terms of age and gender.} sample of} 17 {frequent} users {of RecSys}. 
Based on our findings, we conceptualize experienced serendipity as \textit{a user experience in which the user unintentionally encounters content that feels fortuitous, refreshing, and enriching}.
While these three components are both necessary and sufficient for capturing the concept, they can be fulfilled through various combinations of conditions, allowing for nuanced and flexible interpretations. 

This conceptualization addresses longstanding inconsistencies and contradictions in prior conceptualizations of serendipity, reframing them as distinct flavors of experienced serendipity.  
Furthermore, the clear rules we formulate clearly delineate what experienced serendipity is---and what it is not---laying the groundwork for a validated and standardized operationalization of the concept.

\subsection{Limitations \& Future Work}
Our study is not without limitations. 
First, our sample included only participants aged 20 to 60, excluding two important demographic groups. 
Young people under 20, who are heavy users of digital platforms, likely differ significantly from our sample. 
Similarly, people over 60 years old, though less frequent users, may exhibit unique patterns of engagement with digital platforms that could offer valuable insights.
Second, our participants were exclusively {Belgian.}
This narrow scope may limit the generalizability of our findings. 
As a result, our conceptualization may overlook certain conditions of experienced serendipity or their specific expressions across broader populations.
{In future research, we intend to address this limitation by conducting additional interviews with participants from different geographic and cultural backgrounds.}
{Third, participants primarily discussed their experiences with large, international platforms. 
Future research could explore how well our framework applies to smaller, local platforms.}

Based on the conceptual framework formulated in this paper, we intend to 
develop a validated and standardized questionnaire that captures experienced serendipity.
Since experienced serendipity is inherently subjective---being a user experience---, we believe a questionnaire is the most appropriate instrument for capturing it.
Such a questionnaire will provide a ground truth, forming the foundation for further progress~\cite{kurovski_21_2024}.
Once established, it can serve as a baseline for developing offline metrics by correlating these with questionnaire results~\cite{kurovski_21_2024}.
The conceptualization developed in this paper is an essential precondition for this questionnaire, as it clearly describes what should be operationalized in it.

\section{Ethical \& Human Subjects Considerations}
As per the regulations of our institute, an ethical review was not required for this study. 
Informed consent was asked and given before starting the interviews as mentioned in Paragraph~\ref{para:consent}. 
{As per the data protection guidelines of our institute, we pseudonomyzed the interviews during transcription, deleted the original recordings, and stored the pseudonomyzed transcripts in a secure storage provided by our institute. 
Lastly, we invited participants to review the final manuscript to confirm that their quotes were faithfully translated into English and properly interpreted.}

\begin{acks}
This work was supported by the Research Foundation Flanders (FWO) under grant number S006323N.
We are grateful to all our participants for sharing their valuable time and insights and to our amazing colleagues and reviewers for their thorough reviews and the many helpful suggestions they provided on this paper.
\end{acks}

\bibliographystyle{ACM-Reference-Format}
\bibliography{bibtex}

\end{document}